\documentclass[aps,prl,twocolumn,groupedaddress]{revtex4}

\usepackage{graphicx}
\usepackage{amsfonts}
\usepackage{amssymb}         
\usepackage[squaren,thinspace]{SIunits}
\usepackage{onlyamsmath,color}

\newcommand{\1}{\ensuremath{\langle1\bar{1}0\rangle}}

\newcommand{\dIdU}{\ensuremath{\mathrm{d}I/\mathrm{dU}} }

\newcommand{\etal}{\textit{et al. }}

\begin{document}

\title{Parity effect in ground state localization of antiferromagnetic chains coupled to a ferromagnet}

\author{Simon Holzberger$^1$, Tobias Schuh$^1$,
Stefan Bl\"ugel$^2$,  Samir Lounis$^2$, 
Wulf Wulfhekel$^1$ $^*$}

\affiliation{$^1$ Physikalisches Institut, Karlsruhe Institute of Technology, Wolfgang-Gaede-Strasse 1, 76131 Karlsruhe, Germany \\$^2$Peter Gr\"unberg Institut and Institute for Advanced Simulation, Forschungszentrum J\"ulich \& JARA, 52425 J\"ulich }

\date{\today}

\begin{abstract}
We investigate the ground states of antiferromagnetic Mn nanochains on Ni(110) by spin-polarized scanning tunneling microscopy in combination with theory. 
While the ferrimagnetic linear trimer experimentally shows the predicted collinear classical ground state, no magnetic contrast was observed for dimers and tetramers where non-collinear structures were expected based on \textit{ab-initio} theory. This striking observation can be explained by zero-point energy motion for even numbered chains derived within a classical equation of motion leading to non classical ground states. Thus, depending on the parity of the chain length, the system shows a classical or a quantum behavior.
\end{abstract}

\maketitle


Magnetism is ultimately caused by the spin degree of freedom of the electrons. If phase coherence of the electrons was preserved, the quantum nature of spin would potentially allow to encode quantum information in spintronic devices \cite{Leuenberger2001}. Thus, realizing magnetic quantum devices necessarily involves the understanding of the spin of nanoscopic structures on a quantum mechanical level.
Antiferromagnetic nanostructures are by far not as well studied as their ferromagnetic counterparts. This deficiency lies in the inherent experimental and theoretical difficulties which have to be overcome to understand antiferromagnets. Up to date, even for very simple structures such as the one-dimensional antiferromagnetic chain, the ground state is unknown. While neutron diffraction of one dimensional antiferromagnets often revealed a simple, {\it i.e.} classical, alternative orientations of the spins \cite{Neel} - called the N{\'e}el state - half integer spin chains are for example expected to be in a complex entangled ground state \cite{Balents2010}.
The ground state becomes even more complex when competing exchange interactions exist, leading to magnetic frustration, non-collinear spin structures \cite{Lacroix2010} or to correlated ground states predicted by the Anderson resonating valence bond model \cite{Anderson1956}. 
Geometric frustration of the Heisenberg antiferromagnet on a triangular lattice is the standard example of a magnetically frustrated system \cite{Wannier1950,Wulfhekel2007,Gao2008,Wasniowska2010}. Here we show that antiferromagnetic chains display a classical N\'eel state for odd numbered length and an entangled state for even numbered length when competing exchange interactions and spin-orbit interactions are present. Thus, the chains alternate between opposite nature of the ground states just by the removal or addition of a single atom.

In the present study, i.e. in atomic Mn chains on Ni(110), the frustration arises from the antiferromagnetic coupling within the chain competing with the ferromagnetic coupling of the chain atoms to the substrate. Lounis \etal showed theoretically, that this can lead to an even-odd effect, where the magnetic structure crucially depends on the parity of the number of atoms in the chain \cite{Lounis08}. Sole consideration of the predominant antiferromagnetic coupling within the chain leads to an antiparallel order of the magnetic moments. Therefore, odd-numbered chains exhibit a net magnetic moment, in contrast to even-numbered ones. Switching on the weaker ferromagnetic coupling between the atoms of the chain and the substrate thus acts differently on the two kinds of chains. While odd-numbered ones retain their collinearity and the net moment of the chain aligns with that of the substrate, even-numbered chains develop a more complex ground state. In a presumed collinear state, the total  magnetic exchange energy of the Mn chains to the ferromagnetic substrate is independent of the direction of the Mn moments.  
It can, however, be lowered when a non-collinear spin-structure develops. While magnetic exchange energy has to be paid to tilt the spins of the even-numbered chain from the ideal collinear state, a net spin of the chain develops that points in the direction of the substrate magnetization thus giving rise to an energy gain due to the exchange with the substrate.

Mn chains of lengths up to 6 atoms on a Ni(110) surface were simulated using the Korringa-Kohn-Rostoker Green function method~\cite{SKKR1,SKKR2} as expressed within density functional theory (DFT) taking into account non-collinear spin structures~\cite{Lounis05,Lounis08}. The results were then mapped to a classical Heisenberg model in which magnetic exchange energies between first neighbor atoms were taken into account~\cite{Liechtenstein1987}. As shown before for chains on Ni(001)~\cite{Lounis08}, this model catches the important features observed in the \textit{ab-initio} calculations. 

\begin{figure}
	\centering
		\includegraphics[width=0.7\columnwidth]{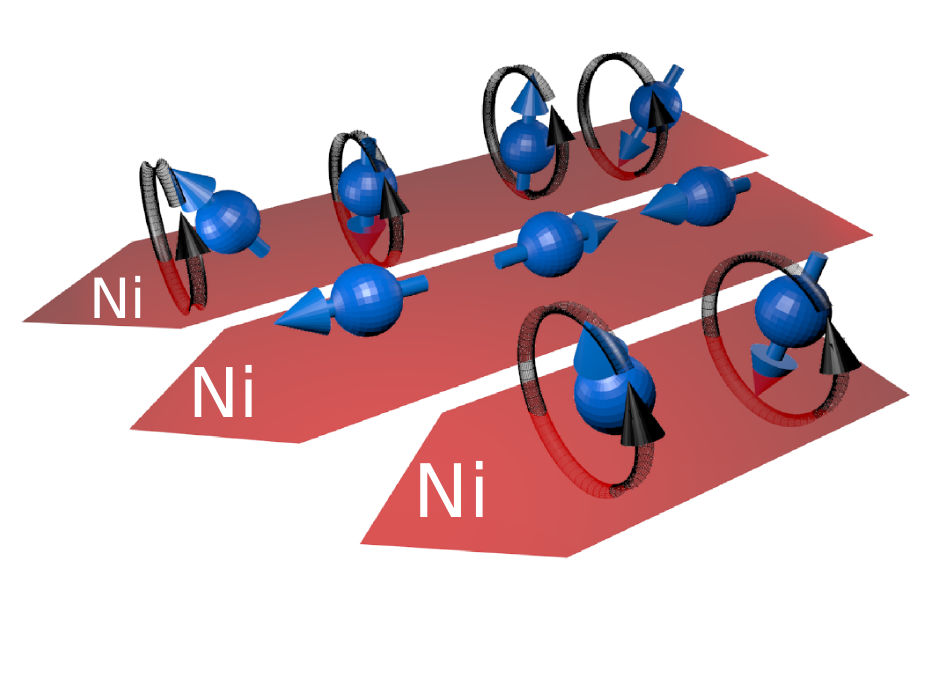}
	\caption{DFT ground states of antiferromagnetically coupled Mn chains on Ni(110). While the linear trimer shows a collinear ferrimagnetic order with magnetic moments (blue vectors) parallel to the Ni magnetization (red arrow), even-numbered chains show a non-collinear magnetic structure. Without spin-orbit interaction, the magnetic moments can be coherently rotated around the Ni magnetization without changing the energy as depicted by the black circles.}
	\label{fig:MnNi110-states}
\end{figure}

Figure \ref{fig:MnNi110-states} illustrates the ground states of linear dimer, trimer and tetramer chains on Ni(110) calculated within this framework. As expected, the trimer chain is in a collinear magnetic state where the net spin of the chain aligned to the substrate magnetization. The even numbered chains show a non-collinear magnetic configuration with moments strongly deviating from the substrate magnetization direction.
Neglecting spin-orbit interaction, the magnetic moments of the chains can be rotated coherently around the direction of magnetization of the substrate without affecting the total energy. 
When spin-orbit interaction is taken into account, the rotational degeneracy is lifted and the Mn moments prefer an out-of-plane orientation, while Ni(110) is magnetized in the surface plane due to shape anisotropy. Thus, two degenerate ground states are predicted. 

So far there has been no experimental report on this effect. 
Currently, only spin-polarized scanning tunneling microscopy (Sp-STM) can reveal antiferromagnetism on the atomic scale \cite{Heinze00,Gao2008a}. Low temperature STM has been used to investigate the quantum nature of small magnetic clusters \cite{Hirjibehedin06,Meier2008,Balashov2009,Zhou2010,Miyamachi2011,Khajetoorians2012}. Furthermore, STM is capable of moving adatoms thus offering the possibility of assembling and probing at the same time \cite{Eigler90}. 
 In this work a home-built STM operating at $\unit{4.2}\kelvin$ and under ultrahigh vacuum conditions was used in combination with W-tips coated with $\unit{10}{ML}$ of Fe,  $\unit{15}{ML}$ of Mn, or $\unit{30}{ML}$ of Co for the spin-polarized measurements. 

\begin{figure}[b]
	\centering
		\includegraphics[width=.7\columnwidth]{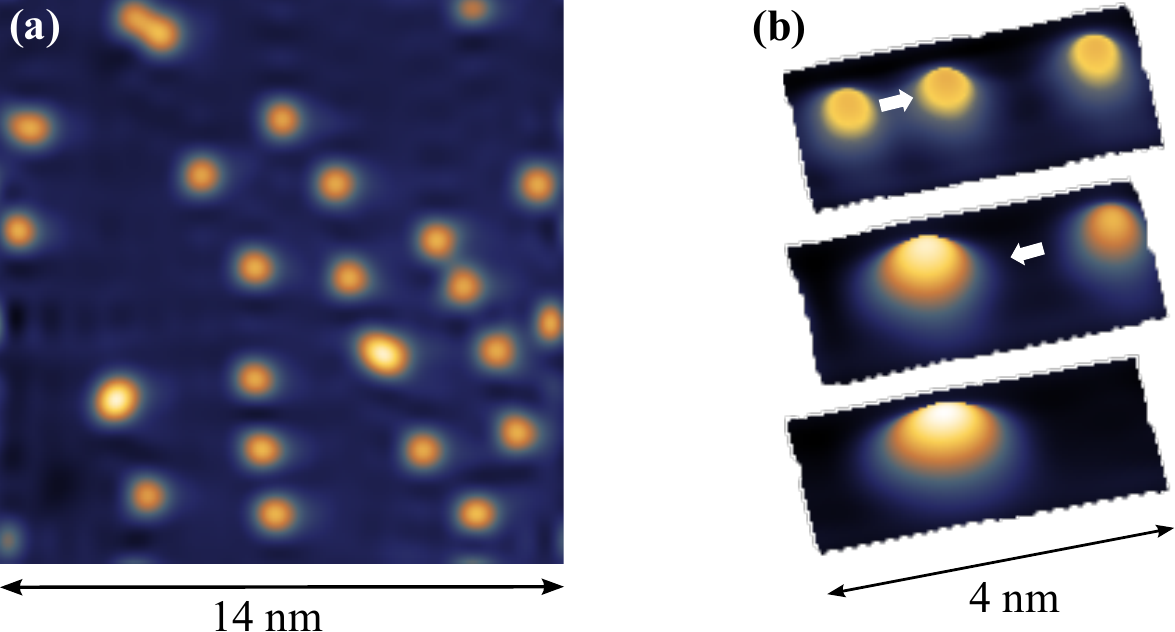}
	\caption{STM images of Mn atoms on Ni(110). (a) Sample with 0.02 ML Mn/Ni(110) deposited at \unit{4.2}{K} showing mainly isolated atoms. (b) Formation of a linear Mn trimer by atomic manipulation.}
	\label{fig:Vgl-deposition}
\end{figure}

\begin{figure}[h]	
	\centering
		\includegraphics[width=.7\columnwidth]{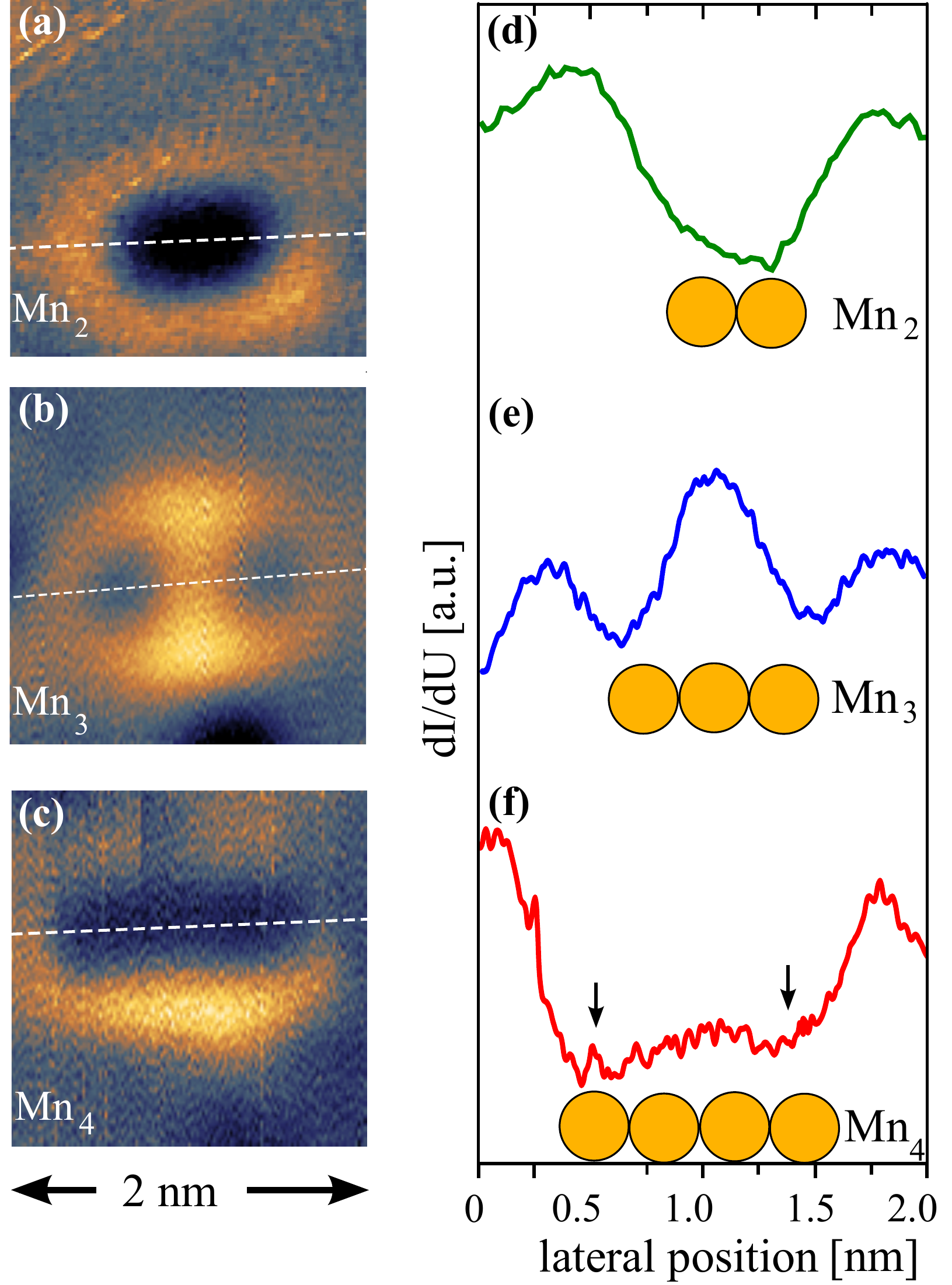}
	\caption{(a-c) Sp-STS results obtained with Fe-coated tip at \unit{350}{mV}, \unit{6}{nA} with the same tip on a dimer, trimer and tetramer, respectively. (d-f) Linescans along the dashed lines in the corresponding images (a-c).}
	\label{fig:sp-STS-results}
\end{figure}

We first deposited  $\unit{0.02}{\text{ML}}$ Mn with the sample held at $\unit{4.2}\kelvin$ showing primarily single Mn adatoms (see Fig.\ref{fig:Vgl-deposition} (a)). Mn chains with the intrinsic nearest-neighbour spacing of $\unit{2.49}\angstrom$ were then assembled by atomic manipulation along the close-packed \1 direction of the substrate (see Fig.  \ref{fig:Vgl-deposition} (b)). 
The assembly was limited to tetramers as longer chains were unstable 
due to the large lattice mismatch between Mn and Ni \cite{CRC}.
Tunneling spectroscopy was used to determine the electronic structure of the chains revealing no resonances below \unit{1}{eV} (see Supplementary material).  
Using Sp-STM with in-plane magnetized Fe coated tips, we investigated the magnetic structure of Mn chains on the atomic scale. 
For trimers, the measurements revealed a strong spin contrast along the chain at a sample bias of \unit{350}{mV}  (see Fig. \ref{fig:sp-STS-results}(b)). Similarly, the line section along the trimer axis displays two minima at the edge adatoms and a maximum in their centers. This is in full agreement with the predictions of antiferromagnetic odd-numbered chains displaying a simple collinear spin structure (see Fig. 1). 
The dimer (Fig. \ref{fig:sp-STS-results} (a)) does not show a contrast in agreement with the predictions.  In the calculated magnetic configuration, the projection of the individual Mn magnetic moments on the substrate magnetization are identical. 
Neither, the tetramer (see Fig. \ref{fig:sp-STS-results} (c)) displays a strong contrast. At a closer look, the line section shows a small depression of the \dIdU signal at the edge atoms of the chain (arrows in Fig. \ref{fig:sp-STS-results} (f)) . This is expected for the tetramer, as the edge atoms align more to the Ni substrate moments. Thus, all chains show the expected projections of the magnetic moment along the substrate magnetization.

Dimers and tetramers were suggested to have a non-collinear magnetic structure and as a consequence, they should display a more complex spin contrast when using a tip magnetized perpendicular to the Ni moments. In experiments with different Mn tips of random magnetic orientation,  Fe tips with arbitrary in-plane orientation or Co tips with out-of-plane orientation, we, however, never observed such a contrast. This rather suggests that the expectation value of the Mn moments perpendicular to the magnetization direction of the substrate does not vary along the even-numbered chains under the experimental conditions. Possibly, spin-fluctuations are present induced by external perturbations, e.g.
temperature or inelastic scattering induced by the tunneling electrons. 

We, however, argue in the following that when treating the energy of the spin configuration within classical equations, we do not predict a classical ground state but a non-classical state due to zero-point motion even at zero temperature \cite{Anderson1956}. The complexity of the magnetic interactions in non-collinear structures with spin-orbit interactions hinders the use of ab-initio methods beyond DFT,  {\it e.g.} a time-dependent procedure  (see for example Ref.~\cite{Lounis10,Lounis11}). Also a model Hamiltonian of a quantum Heisenberg system with localized spins and exact diagonalization \cite{Hirjibehedin06} describing effects like magnetization tunneling \cite{Wernsdorfer1999} cannot be used, as it neglects the itinerant nature of the system.
Instead, we take a new pathway to describe even-numbered chains following ideas of Leggett \textit{et al.} \cite{Leggett1987} in the framework of the "spin-boson" problem. By integrating out the electronic degrees of freedom described within DFT, we compute the interatomic exchange interaction and the intraatomic spin-orbit interaction. The parameters extracted from DFT are then fed into classical equations of motion. The resulting equations are then treated in the limit of quantum mechanics to investigate the spin-dynamics of the system. 

The equation of motion \cite{Gilbert2004} for a magnetic atom without damping is given by 
\begin{equation}
\frac{\partial \textbf{M}}{\partial t}= -\gamma \textbf{M} \times \textbf{H}_{\mathrm{eff}}, \label{eq:Landau}
\end{equation}
where $\textbf{H}_{\mathrm{eff}}$ is the effective magnetic field acting on the magnetic moment $\textbf{M}$, and $\gamma$ is the gyromagnetic ratio. $ \textbf{H}_{\mathrm{eff}}$ can be determined from the total energy according to
\begin{equation} \textbf{H}_{\mathrm{eff}}=\frac{-\nabla_\textbf{M} E}{\gamma},
\end{equation}
where $E$ is the total energy corresponding to a Heisenberg Hamiltonian which includes magnetic anisotropy and in which the magnetic exchange energies between first neighbors are considered. For the Mn dimer we find: 
\begin{equation}
E(\theta,\phi)=-J_1 \cos(2\theta)-2J_2 \cos(\theta)+ 2K \sin^2(\theta) \cos^2(\phi)
\end{equation}
Here, $\theta$ is the polar angle of Mn magnetization to the magnetization direction of the Ni substrate and $\phi$ is the azimuthal angle of the magnetization direction of the Mn atoms with respect to the easy axis. The Ni moments were treated as rigid. $J_1$ and $J_2$ are the exchange constants for Mn-Mn and Ni-Mn exchange, respectively. $K$ is the uniaxial anisotropy per adatom of the Mn dimer for rotation around the Ni magnetization direction. 
The equation describes the coupled dynamics of $\theta$ and $\phi$ in a harmonic potential for small deviations from the ground state with $\phi_0$=0$^\circ$, 180$^\circ$.

From our \textit{ab-initio} calculations we determine $J_1=\unit{-221.4}{meV}$, $J_2=\unit{116.8}{meV}$ and  $K=\unit{-0.3}{meV}$  which result in $\theta_0=75^{\circ}$ and an easy direction of the Mn moments along the surface normal.
The energy barrier to coherently and adiabatically rotate the moments of the Mn atoms is proportional to the anisotropy barrier $K$ multiplied by $2\sin^2(\theta_0)=\unit{0.56}{meV}$. When solving the coupled equation of motion in $\theta$ and $\phi$ around the ground state, we obtain an eigen-frequency $\omega$ for the precession of the magnetic moments. This precession involves both $\phi$ and $\theta$ (see {\it e.g.} Ref.~\cite{Valstyn62}). It corresponds to a periodic oscillation of the two variables in a local well of the potential. It can be treated to lowest order as a harmonic oscillator. This frequency defines thus a zero-point fluctuation energy $E_{o}$:
\begin{widetext}
\begin{eqnarray}
 &E_{o}=\frac{1}{2} \hbar \omega = \\
 &\frac{g\mu_B}{2M}
\sqrt{-4K\cos(2\phi_0)(4J_1\cos(2\theta_0)+2J_2\cos(\theta_0)+4K\cos(2\theta_0)\cos^2(\phi_0))-(4K\cos(\theta_0)\sin(2\phi_0))^2} \nonumber
\end{eqnarray}
\end{widetext}
where $g \approx 2$ is the g-factor and $\mu_B$ the Bohr magneton. Surprisingly, also the exchange interactions enter the zero-point fluctuation energy leading to $E_{o}$ of the order of \unit{8.9}{meV}. Thus, in the ground state, the magnetic moments have a much larger fluctuation energy than the anisotropy barrier and the system can overcome the barrier. 

Similar equations for the tetramer also reveal, that the zero-point motion is large enough to overcome the anisotropy barrier (see Supplementary material). In the framework of the "spin-boson" problem, this case represents the situation where the matrix element for tunneling between the two localized states (mediated by the exchange interaction) is larger than the barrier (mediated by the spin-orbit interaction). In this pathological case of the "spin-boson" problem, the states do not localize in one or the other classical state even at $T=\unit{0}{K}$ \cite{Leggett1987}. It is crucial and very instructive to analyze the previous equation: 
The barrier height and $E_o$ depend on both the anisotropy and the exchange energies. While the maximal value of the barrier height is limited by the anisotropy, the zero-point energy strongly depends on the exchange constants. 
One notices that for typical values of the exchange being much larger than the anisotropy, the zero-point energy is also much larger than the anisotropy. 
By decreasing $J_1$ and omitting $J_2$ (paramagnetic substrate), the zero-point energy decreases quickly. Thus at the limit of weak interactions between the adatoms and a finite barrier due to anisotropy, the system exhibits a localized, {\it i.e.} N\'eel,  ground state \cite{Loth2012}.

\begin{figure}[b]
	\centering
		\includegraphics[width=.8\columnwidth]{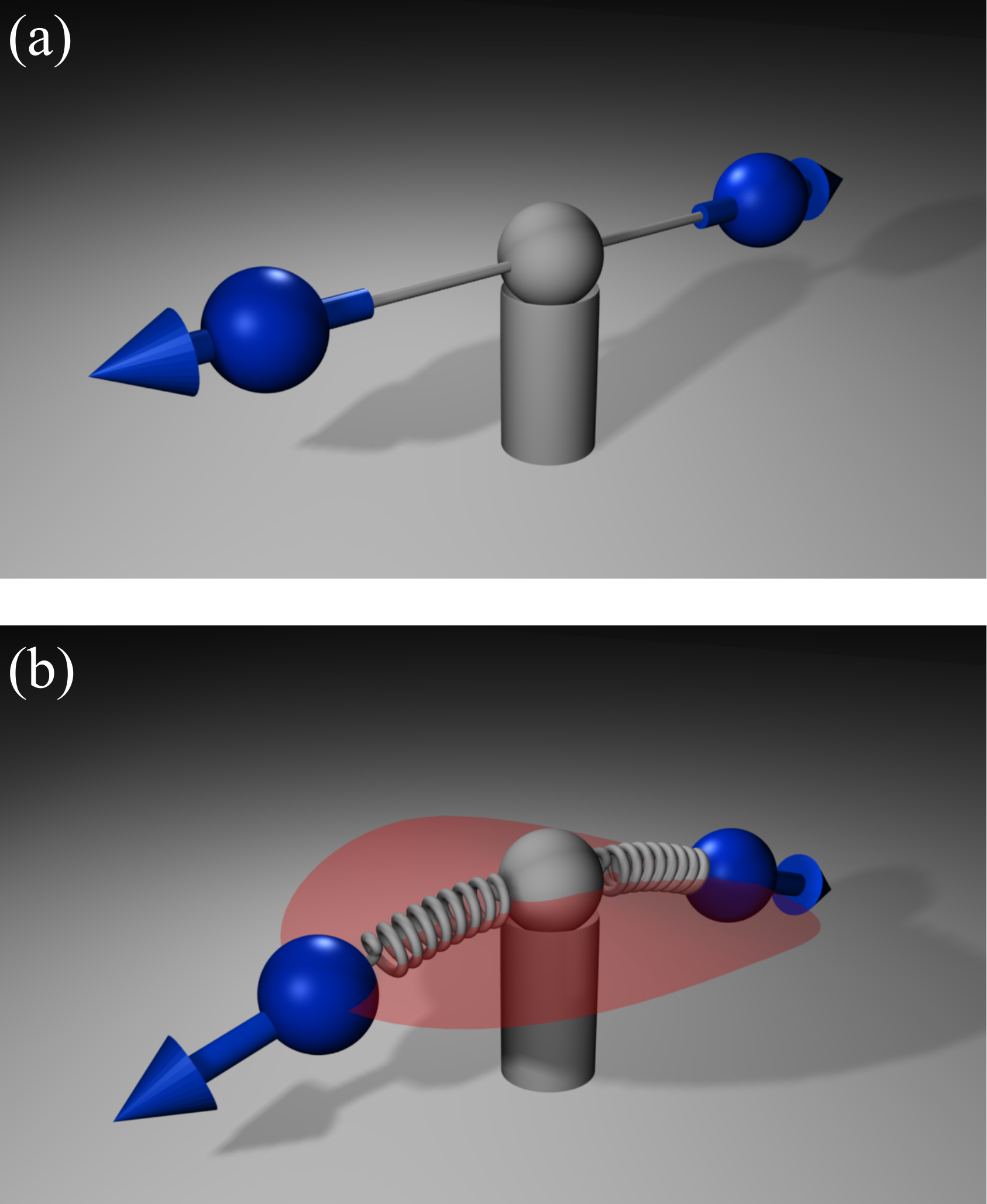}
	\caption{ Classical analogon of the dimer ground state (a) in the limit of an infinite antiferromagnetic exchange of the Mn dimers and in the absence of spin-orbit interaction and (b) when the antiferromagnetic exchange is finite in the presence of a magnetic anisotropy.}
	\label{fig:classic}
\end{figure}

To illustrate the equations of motion, we map the problem onto a mechanical analogon. In the limit of infinite Mn exchange and absence of spin-orbit interaction all antiferromagnetic Mn configurations irrespective of their orientation to the substrate magnetization direction are degenerate. The mechanical analogon would be two spinning masses (spinning axis indicated by the blue arrows representing the magnetization direction) coupled by a rigid stick (see Fig. \ref{fig:classic} a). The exchange interaction to the substrate creates an effective field equivalent to gravity pulling the two masses down.
The ground state is, however, continuously degenerate without a barrier to rotate the spins. Thus in the quantum limit of the mechanical model, the dimer can be in an arbitrary superposition of different states leading to a vanishing expectation value of the magnetic moment for the individual atom.
If we assume a finite antiferromagnetic exchange, the dimer deviates from a strictly antiparallel orientation in the exchange field of the ferromagnet much alike in the mechanical model, when the two masses are connected with springs (see Fig. \ref{fig:classic}b).  When taking into account uniaxial magnetic anisotropy, two degenerate ground states evolve as indicated in Fig. \ref{fig:classic}b separated by a shallow potential barrier, shown in red. These ground states represent the degenerate states as determined with DFT with spin-orbit interaction. As the vibration of the spring is coupled to a rotation of the dimer magnetization due to precession, we expect that the mechanical model populates a discrete ground state at $T=\unit{0}{K}$, in case the barrier between the two ground states cannot be overcome by zero-point fluctuations of the spring \cite{Leggett1987}. If it can, a superposition of the two degenerate ground states is the true ground state, as in our case due to the large zero-point energy caused by the strong Mn-Mn exchange.

In conclusion, we have shown that although there is a finite magnetic anisotropy energy acting as a potential barrier between the degenerate DFT ground states in even-numbered chains, a zero-point energy provides a mean for fluctuations between the two degenerate states. The latter one is found to be surprisingly large and should be considered when describing antiferromagnetic nano-objects. 
Just by adding or removing one atom of the chain - changing parity - the system changes its magnetic behavior completely and behaves classically. This is due to the net spin of the chain coupling to the macroscopic magnetization of the Ni substrate.
We believe that our findings can show a path to create magnetically stable antiferromagnetic structures, {\it i.e.} to raise the blocking temperature.
This intriguing result is obtained without requiring a quantum Heisenberg model but by treating the equation of motion for the magnetic moments as a quantum equation. In this respect, we believe that this approach is general and could lead to a better understanding of the dynamics of small spin systems.


{\bf Acknowledgements}

S.L. acknowledges discussions with P. H. Dederichs and the support of the HGF-YIG Programme VH-NG-717 (Functional nanoscale structure and probe simulation laboratory-Funsilab). W.W. acknowledges discussions with J. Schmalian and funding by the Deutsche Forschungsgemeinschaft (DFG grant WU 349/4-1).


\end{document}


\title{Supplementary Information: Parity effect in ground state localisation of antiferromagnetic chains coupled to a ferromagnet}

\author{Simon Holzberger$^1$, Tobias Schuh$^1$,
Stefan Bl\"ugel$^2$,  Samir Lounis$^2$, 
Wulf Wulfhekel$^1$}

\affiliation{$^1$ Physikalisches Institut, Karlsruhe Institute of Technology, Wolfgang-Gaede-Strasse 1, 76131 Karlsruhe, Germany \\$^2$Peter Gr\"unberg Institut and Institute for Advanced Simulation, Forschungszentrum J\"ulich \& JARA, 52425 J\"ulich }

\maketitle
\subsection*{Sample preparation}

All experiments have been performed in ultra high vacuum ($p<10^{-10}$ mbar).
The Ni(110) crystal was cleaned by repeated cycles of Ar$^+$ ion sputtering ($\unit{1.5}{\kilo\volt}$, $\unit{30}\minute$) and annealing to $\unit{970}\kelvin$ until no contaminations were found by low energy electron diffraction (LEED). Atomically resolved scanning tunneling microscope (STM) images revealed a clean surface with terrace sizes exceeding $\unit{200}{\nano\meter}$ in width.

Mn was evaporated \textit{in situ} from high purity Mn grains placed into a Mo-crucible heated by electron bombardment with typical deposition rates of approximately $\unit{0.1}{\text{ML}\per\minute}$. For Mn deposition above room temperature the formation of an ordered surface alloy at half monolayer coverages of Mn on Ni(110) has been deduced from LEED experiments \cite{Rader01, deSantis04}. This phase is characterized by a c$(2\times2)$ surface unit cell. Thus, deposition below room temperature is indispensable. When keeping the sample at $\sim\unit{170}\kelvin$ during the deposition process, no superstructure was observed by LEED. This indicates that there is no ordered intermixing between manganese and nickel. 

Although for low coverages of Mn ($\sim\unit{0.02}{\text{ML}}$) one observes close-packed monoatomic chains of adatoms over a wide range of temperatures, the formation of chains was always accompanied by isolated protrusions in the nickel substrate as seen by STM. We attribute these protrusions (apparent height $\unit{30}{\pico\meter}$) to incorporated manganese atoms, as their density roughly scaled with the amount of deposited Mn and is reduced for lower deposition temperatures. This was further confirmed by comparing scanning tunneling spectroscopy (STS) measurements performed on protrusions in the substrate and on Mn adatoms exhibiting a similar structure. This, in turn, implies that such chains of adatoms are intermixed Mn/Ni chains. Therefore, direct deposition of Mn on the sample held at cryogenic temperatures ($\unit{4}{K}$) and atomic manipulation is necessary for pure Mn chains.

As STM tips, chemically etched W-wires were used that were cleaned \textit{in situ} by Ar$^+$ ion sputtering and heating to \unit{3700}{K}. For Sp-STS measurements the cleaned W-tips were coated with $\unit{10}{ML}$ of Fe or $\unit{15}{ML}$ of Mn or $\unit{30}{ML}$ of Co and gently annealed. 

The STM was cooled to \unit{4.2}{K} using a liquid He bath cryostat. Mn chains were assembled by atomic manipulation, i.e., by positioning the tip on a desired atom, reducing the resistance of the tunneling gap to 40 k$\Omega$ and moving the tip along the groves of the Ni(110) surface. When returning to the normal tunneling conditions of gap resistances larger than 1 M$\Omega$, the success of the lateral manipulation was checked by imaging. Sp-STM was carried out with coated tips and modulating the bias voltage by \unit{10}{mV} at high frequencies (\unit{10}{kHz} range). The differential conductance was then recorded as function of tip position using a lock-in detection.

\subsection*{Confined electronic states}
\begin{figure} [hb]
\centering
\includegraphics[width=.8\columnwidth]{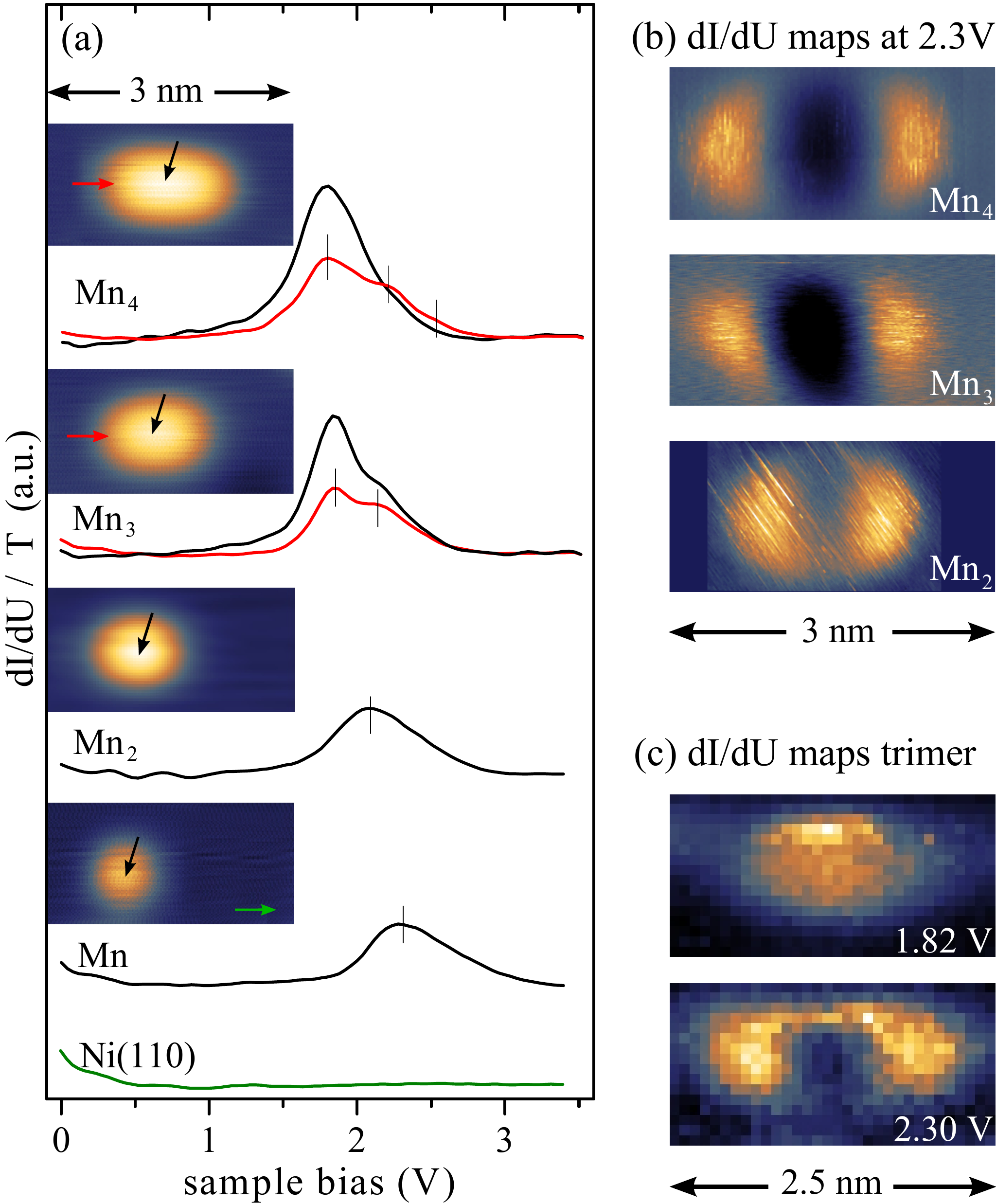}
\caption{\label{fig:LDOS}(a) Normalized \dIdU spectra taken on Mn nanochains on Ni(110) acquired at the indicated positions in the corresponding topographic image. (b) \dIdU maps of a dimer, trimer, and tetramer at \unit{2.3}{V}, showing that this state is predominantly located at the edges of the chains. (c) The \dIdU maps of a trimer taken at \unit{1.82}{V} and at \unit{2.3}{V} show that the states that correspond to the peaks in the \dIdU spectra have different spatial distributions.}
\end{figure}

In STS  measurements, the differential conductance \dIdU is proportional both to the local density of states (LDOS) of the sample and the tunneling matrix element \cite{Tersoff1983}. By means of a proper normalization of the \dIdU signal, the LDOS of the sample can be inferred \cite{Ukraintsev96}. Fig. \ref{fig:LDOS}~(a) shows the normalized differential conductivities measured at the indicated positions of the corresponding Mn chains and on the Ni substrate. As the projected electronic structure of Ni(110) does not exhibit a pseudo gap, the retrieved LDOS is rather constant.

In contrast, all spectra taken on Mn chains show distinct features. 
For a single Mn atom, there is a broad peak at $\unit{2.4}\volt$ (full width at half maximum\,\,$\sim\unit{0.8}\electronvolt$). Spectra taken at the center of trimers and tetramers have a characteristic substructure separable into several overlapping peaks. Changing the lateral position of acquisition leads to a change in the relative intensity of the sub-peaks but leaves their energy unchanged. Maps of the local differential conductance at fixed bias voltage show a standing wave pattern (see \ref{fig:LDOS} (b) and (c)). The peaks in the \dIdU spectra arise from electronic states confined to the Mn chains. Remarkably, the onset of the lowest peak in energy shifts to lower bias voltages for longer chains (from $\unit{2.4}\volt$ for the single atom to $\unit{1.7}\volt$ for the tetramer). This is in qualitative agreement with a simple particle in a box model.

The results for atomic Mn chains on Ni(110) are consistent with similar results previously reported for copper chains on Cu(111) \cite{Folsch2004} and for gold chains on NiAl(110) \cite{Nilius02}. Most importantly, no standing wave pattern has been observed at low sample bias ($<\unit{1}\volt$) indicating a homogeneous LDOS at these energies.


\subsection*{Ab-initio calculations}

The \textit{ab-initio} calculations were performed using density functional theory as implemented in the full-potential Korringa-Kohn-Rostoker Green function method \cite{SKKR1}, where the non-collinear treatment was used~\cite{Lounis05} including relativistic effects~\cite{SKKR2}. The method is ideal for investigating nanostructures on substrates in a real space embedding approach. A Dyson equation is solved where the Green function of interest $G$ is related to the Green function of the perfect substrate $G_0$ via the change of potential $V$ induced by the perturbing nanostructure: $G = G_0 + G_0VG$ (in a matrix notation). From the green function $G$, the charge and spin densities are extracted self-consistently.

The magnetic exchange interactions $J_{ij}$ mentioned in the paper and used in the Heisenberg model as well as in the formula for the zero-point energy describe the pairwise exchange interactions within the Mn nano-wires. Their values were calculated considering infinitesimal rotation of the magnetic moments~\cite{Liechtenstein1987}. This methods gives a good account of the interactions since the results obtained with the Heisenberg model are very close to the first-principles calculations as shown in Ref.~\cite{Lounis08}. The results are shown in Table~\ref{Jij}. One notices that the total length of the chain has a minor influence on the nearest neighbor exchange constants ($J_\textup{n.n.}\sim\unit{-200}{meV}$), whereas there is a dramatic decrease with increasing distance. All exchange constants between second nearest neighboring adatoms are positive, which explains that these adatoms have moments oriented along the same directions.

\begin{table}
[ht]

\begin{center}
\caption{\label{Jij} Exchange constants in units of meV between Mn adatoms within different nano-wires.}
\begin{ruledtabular}
\begin{tabular}{lccccc}
 $J_{ij}$         & Dimer  & Trimer  &  Tetramer         & Pentamer & Sextamer\\
\hline
 {12}   & -221.43& -207.13 & -179.31   & -187.04 & -191.51 \\
 {13}   & /      & +31.50  & +19.14   & +11.76 & +11.58 \\
 {14}   & /      &   /     & -8.77   & -7.17 & -5.05 \\
 {15}   & /      &   /     &  /      & +2.4 & +3.26\\
 {16}   & /      &   /     &  /      &  /   & -0.57 \\
 {23}   & /      & -207.13 & -191.00 & -172.63 & -180.62\\
 {24}   & /      &   /     & +19.14 &  10.12 &  +3.73\\
 {25}   & /      &   /     & /      &  -7.17 & -5.55\\
 {26}   & /      &   /     & /      &    /   & +3.26\\
 {34}   & /      &   /     & -179.31&  -172.61 & -159.03\\
 {35}   & /      &   /     & /      &  +11.78 & +3.26\\
 {36}   & /      &   /     & /      &   /      & -5.05\\
 {45}   & /      &   /     & /      &  -187.04 & -180.62\\
 {46}   & /      &   /     & /      &  /       &  +11.58\\
 {56}   & /      &   /     & /      &  /       &  -191.51\\ 
\end{tabular}
\end{ruledtabular}
\end{center}
\end{table}

In Table~\ref{table:noncol}, we provide details on the rotation angles and the magnetic moments of the non-collinear states of the even-numbered chains. In addition, we calculated the energy differences $\Delta E$ between the non-collinear state and the ferrimagnetic solution for even-numbered chains, showing that the non-collinear states are energetically favored.
 
\begin{table}[ht]
\begin{center}
\caption{\label{table:noncol} {\em Ab initio} results for 
the nanochains: size and angle of the magnetic moments
as well as total energy differences $\Delta E$ between the non-collinear and antiferromagnetic solutions.
In every pair of wire atoms connected by a ``-'' sign, the  azimuthal 
angles $\phi$ are equal to 0$^o$ and 180$^o$ respectively, while the 
magnetic moments and rotation angles $\theta$ are the same.}
\begin{ruledtabular}
\begin{tabular}{lcccc}
 Length           & Adatom             & $\theta $($^o$) &  $M$ ($\mu_B$) & 
$\Delta E$\\
 (adatoms)        &                    &                 &              & (meV/adatom)\\
\hline
 2                &1-2                 &  77             &   3.53       & $-$2.26\\
                  &                    &                 &              &         \\
 4                &1-4,2-3             &  67, 85         &   3.52, 3.29 & $-$5.59 \\
                  &                    &                 &              &         \\
 6                &1-6, 2-5            &  57, 94,       &   3.51, 3.30 & $-$6.33 \\
                  &3-4                 &       76       &         3.27 &          \\
\end{tabular}
\end{ruledtabular}
\end{center}
\end{table}

\subsection*{Zero-point energy of a tetramer}

The total magnetic energy of a tetramer considering only next neighbor exchange interactions and uniaxial anisotropy is given by:
\begin{eqnarray}
&E_4(\theta_1, \theta_2, \phi_1,\phi_2)\nonumber\\
&=-2J_{12} \cos(\theta_1+\theta_2) -J_{23} \cos(2\theta_2)\nonumber\\
&-2 J_{f1} \cos(\theta_1)-2 J_{f2} \cos(\theta_2) \nonumber\\
&+2K( \sin^2(\theta_1)\cos^2(\phi_2)+\sin^2(\theta_2)\cos^2(\phi_2)).\label{eq:ground4}
\end{eqnarray}
where $J_{12}$ and $J_{23}$ are the exchange constants given in Table \ref{Jij}. $J_{f1}$ and $J_{f2}$ account for the ferromagnetic exchange between the substrate and the inner and outer Mn atoms, respectively. From the DFT calculations we obtain the values {\color{black} $J_{f1}=\unit{90}{meV}$} and $J_{f2}=\unit{94}{meV}$. The magnetic anisotropy energy $K$ per atom is taken from the value of the dimer. The orientation of the magnetic moment of the inner atoms is described by $\theta_1$, $\phi_1$ and by $\theta_2$, $\phi_2$ for the outer atoms. The angles in the ground state minimize the energy given in equation \ref{eq:ground4} and are found to be:
\begin{subequations}
\begin{eqnarray}
\theta_{10}&=65^\circ\quad\theta_{20}&=88^\circ\nonumber\\
\phi_{10}&=0^\circ\quad\phi_{20}&=180^\circ\nonumber
\end{eqnarray}
\end{subequations}
{\color{black} where the additional subscript $0$ stands for ground state.}
Note that this approach leads to results {\color{black} similar to} our DFT calculations. The zero-point energy of the tetramer $E_{04}$ is found to be \cite{Lindner2003}:

\begin{eqnarray}
E_{04}&=& {\color{black} \frac{1}{2}}\hbar \omega= {\color{black}\frac{g\mu_B}{2}} \sqrt{\frac{b}{2}\pm\sqrt{\frac{b^2}{4}-c^2}}\label{eq:lsg4}\\
\mathrm{with} \nonumber \\
b&=&\frac{E_{\theta_1\theta_1} E_{\phi_1\phi_1}}{M_1^2\sin^2(\theta_{10})}+\frac{E_{\theta_2\theta_2} E_{\phi_2\phi_2}}{M_2^2\sin^2(\theta_{20})} \nonumber\\
c&=&\frac{E_{\theta_1\theta_1}E_{\theta_2\theta_2} E_{\phi_1\phi_1} E_{\phi_2\phi_2}-E_{\theta_1\theta_2}^2 E_{\phi_1\phi_1} E_{\phi_2\phi_2}}{M_1^2 M_2^2\sin^2(\theta_{10})\sin^2(\theta_{20})} \nonumber
\end{eqnarray}
where $g\approx 2$ is the g-factor and $\mu_B$ the Bohr magneton.
$E_{\alpha\beta}$ is the second derivative of $E_4$ with respect to $\alpha$ and $\beta$ evaluated at the optimum angles, i.\,e., 
\begin{equation}
E_{\alpha\beta}\equiv\left.\frac{\mathrm{d^2}E}{\mathrm{d}\alpha \mathrm{d}\beta}\right|_{\theta_{10}\theta_{20}\phi_{10}\phi_{20}}.\nonumber
\end{equation} 

As $E_{\theta_2\phi_2}=E_{\theta_1\phi_1}=0$ for $\phi_1=\phi_{10}$ and $\phi_2=\phi_{20}$, this derivative does not explicitly appear in \eqref{eq:lsg4}. $M_1$ and $M_2$ are the magnetic moments obtained from the DFT calculations. The lowest possible zero-point energy is thus given by
\begin{equation}
E_{04}=\unit{5.2}{meV},
\end{equation}
which is again larger than the barrier for the rotation in the $\phi$-direction, given by $|2K|( \sin^2(\theta_1)+\sin^2(\theta_2))=\unit{1.1}{meV}$. 



%